\documentclass[10pt,conference]{IEEEtran}
\IEEEoverridecommandlockouts
\usepackage{cite}
\usepackage{amsmath,amssymb,amsfonts}
\usepackage{algorithmic}
\usepackage{graphicx}
\usepackage{textcomp}
\usepackage{xcolor}
\usepackage{capt-of}
\usepackage[ruled,linesnumbered]{algorithm2e}
\usepackage{import}
\usepackage{tikz}
\usepackage{moresize}
\usepackage{enumerate}
\usepackage{listings}
\usepackage{booktabs}
\usepackage{multirow}
\usepackage{colortbl}
\usepackage{array}
\usepackage[inline]{enumitem}
\usepackage[many]{tcolorbox}
\usepackage{soul,xcolor}
\setstcolor{red}
\usepackage{csquotes}
\usepackage{tabularx}
\usetikzlibrary{shapes.misc,decorations.pathreplacing,calligraphy,tikzmark}
\definecolor{Green}{HTML}{3fc380}
\usepackage[htt]{hyphenat}

\newtcolorbox{rqbox}{
    colback=black!15!white,
    colframe=white,
    colbacktitle=white,
    coltitle=black,
    boxrule=1pt,
  sharpish corners,
  boxrule = 0pt,
    left=1mm,
    right=1mm,
    bottom=1mm,
    top=1mm,
  enhanced,
}

\def\BibTeX{{\rm B\kern-.05em{\sc i\kern-.025em b}\kern-.08em
    T\kern-.1667em\lower.7ex\hbox{E}\kern-.125emX}}

\newcolumntype{P}[1]{>{\centering\arraybackslash}p{#1}}

\newcommand{\code}[1]{\texttt{\small#1}}

\newcommand{\approach}[0]{UTGen }

\newcommand{\nCUTS}[0]{346 }
\newcommand{\nPar}[0]{32 }

\newcommand{\approachNS}[0]{UTGen}

\newcommand{\circled}[1]{\tikz[baseline=(char.base)]{
            \node[shape=circle,draw,inner sep=1pt] (char) {#1};}}

\definecolor{commentbgcolor}{rgb}{0.8, 0.9, 0.9} 

\newcommand*\circledbeta{%
  \tikz[baseline=(char.base)]{%
    \node[shape=circle,draw,inner sep=1pt] (char) {$\beta$};}%
}
\newcommand*\circledalpha{%
  \tikz[baseline=(char.base)]{%
    \node[shape=circle,draw,inner sep=1pt] (char) {$\alpha$};}%
}

\definecolor{color1}{HTML}{FFE26F}
\definecolor{color2}{HTML}{0CA789}
\definecolor{color3}{HTML}{E74C3C}

\begin{document}

\bstctlcite{IEEEexample:BSTcontrol}

\title{Leveraging Large Language Models for Enhancing the Understandability of Generated Unit Tests}


\author{\IEEEauthorblockN{Amirhossein Deljouyi,
Roham Koohestani, Maliheh Izadi,
Andy Zaidman}
\IEEEauthorblockA{\textit{Delft University of Technology}\\
Delft, The Netherlands \\
a.deljouyi@tudelft.nl, r.koohestani@student.tudelft.nl, \{m.izadi, a.e.zaidman\}@tudelft.nl}
}


\maketitle


%

\begin{abstract}
Automated unit test generators, particularly search-based software testing tools like EvoSuite, are capable of generating tests with high coverage. 
Although these generators alleviate the burden of writing unit tests, they often pose challenges for software engineers in terms of understanding the generated tests.
To address this, we introduce \textit{\approachNS{}}, which combines search-based software testing and large language models to enhance the understandability of automatically generated test cases.
We achieve this enhancement through contextualizing test data, improving identifier naming, and adding descriptive comments.
Through a controlled experiment with 32 participants
from both academia and industry,
we investigate how the understandability of unit tests affects a software engineer's ability to perform bug-fixing tasks. We selected bug-fixing to simulate a real-world scenario that emphasizes the importance of understandable test cases. 
We observe that participants working on assignments with \approachNS{} test cases fix up to 33\% more bugs and use up to 20\% less time when compared to baseline test cases.
From the post-test questionnaire, we gathered that participants found that enhanced test names, test data, and variable names improved their bug-fixing process. 

\end{abstract}


\section{Introduction}
In today's software-dominated world, 
software reliability and correctness are very important~\cite{KoCHASE2014}. Consequently, automated testing in the form of unit tests has become a crucial element for software engineers in ensuring high-quality software~\cite{beck2003test-driven,khatamiSPE2024,khatamiSCAM2023}.
Despite the widely acknowledged importance of testing, writing tests is tedious and time-consuming~\cite{beller2019developer,bellerFSE2015,bellerICSE2015,aniche2022how-developers}.
To alleviate this burden on developers and testers, 
the research community has devoted considerable effort 
on investigating automatic test generation approaches~\cite{ali2010a-systematic,baresi2010testful:,fraser2011evosuite,fraser2015does,derakhshanfarTSE2023,brandtTSE2024}. 
Among the notable test generators are Randoop~\cite{Pacheco_2007} and EvoSuite~\cite{fraser2011evosuite}. EvoSuite, for example, is a search-based test generator that employs genetic algorithms to construct a test suite~\cite{FraserEMSE2015} and has demonstrated good results in terms of coverage~\cite{Fraser2013IEEE, Panichella2018AutomatedTC}. 

However, based on insights obtained through industrial case studies, 
there are limitations in terms of the quality of the generated test cases~\cite{ArcuriEMSE2018, Palomba_2016, Palomba2016SBST, GRANO2019, Fraser_2013, Shamshiri_2015, Almasi_2017}.
One critical limitation revolves around the understandability of generated test cases, which involves various aspects such as meaningful test data, proper assertions, well-defined mock objects, descriptive identifiers and test names, as well as informative comments.
Additionally, the difficulty in following the scenario depicted in the test case and the ambiguity surrounding test data significantly hamper clarity~\cite{Almasi_2017,DBLP:journals/ese/BrandtZ22}.

\begin{figure}
\centering
\includegraphics[width=\linewidth]{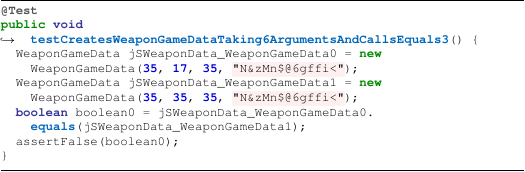}
\vspace{-7mm}
\caption{Motivating Example}
\label{listing:motivating}
\vspace{-6mm}
\end{figure}

Figure~\ref{listing:motivating} provides an example of an EvoSuite-generated test case.
This test case checks the \texttt{equals} method with two objects of \texttt{weaponGameData} with different minimum damage values. 
Here, we see several comprehension challenges:
\begin{enumerate*}
    \item the purpose and functionality of a test method named with five arguments and ``callsEquals3'' is obscure, 
    \item the rationale behind the chosen test data remains unclear,
    \item the identifiers are not providing any additional information, and
    \item the absence of comments leaves the test case without essential explanatory context.
\end{enumerate*}

To address these issues, we aim to enhance automatically generated test cases by focusing on contextual test data, clear test method and identifier names, and adding descriptive comments.
In this study, we investigate the synergy of Search-Based Software Testing (SBST) and Large Language Models (LLMs). While Natural Language Processing (NLP) techniques have shown promise in text generation and optimization~\cite{Zhang_2016, roy2020deeptc}, and LLMs have advanced text-based capabilities~\cite{10329992, yu2023llm, mastropaolo2021icse, Liventsev_2023, wang2023software}, their impact in generating high-coverage test cases for complex systems remains limited~\cite{elhajiAST2024,siddiq2024using}. Conversely, SBST, while effective in coverage, often falls short in test case understandability. 

Our approach, \emph{UTGen}, integrates an LLM into the SBST test generation process.
We hypothesize that this combined approach can leverage the strengths of both techniques to generate effective and understandable test cases.
Our study is steered by three Research Questions (RQs) that consider the effectiveness of the \approach approach, and the understandability of the generated test cases.
\begin{description}
\item[\textbf{RQ$_{1}$}] \textit{Does \approach have the capability to generate effective unit tests by utilizing a combination of LLMs and SBST?}
\end{description}
\vspace{-0.5mm}
The investigation into the effectiveness of the approach seeks to establish whether the non-determinism of both the SBST and LLM components 
impact the ability to generate compilable and high-coverage unit tests.

\begin{description}
\item[\textbf{RQ$_{2}$}] \textit{What is the impact of LLM-improved unit tests' understandability on the efficiency of bug fixing by developers?} 
\end{description}
\vspace{-0.5mm}
When it comes to the understandability of generated test cases, we intend to measure understandability through the ease by which software engineers can perform bug-fixing tasks involving failing test cases, a setup previously used by Panichella et al.~\cite{panichella2016impact}.

\begin{description}
\item[\textbf{RQ$_{3}$}] \textit{Which elements of UTGen affect the understandability of the generated unit tests?}
\end{description}
\vspace{-0.5mm}
We frame RQ$_{3}$ to obtain a deeper understanding about which elements of the \approach approach determine the understandability of the generated test cases. 


The key contributions of our paper are outlined as follows:
\begin{itemize}
    \item \approachNS, our novel approach that integrates an LLM into the SBST process to enhance the understandability of generated unit tests.
    \item The application of \approach on \nCUTS classes to examine the effectiveness of the generated unit tests.
    \item A controlled experiment and a post-test questionnaire with \nPar participants from industry and academia were meant to evaluate the impact of LLM-improved test cases in terms of understandability in a bug-fixing scenario. 
    \item We release a replication package that is publicly available with our implementation, as well as detailed data and results from our evaluation~\cite{replicationpackage}.
\end{itemize}

\section{Background}
\subsection{Search-Based Software Testing}
Automated test generation approaches have been developed in order to reduce testing effort.
Tools like EvoSuite~\cite{fraser2011evosuite} and Randoop~\cite{Pacheco_2007} generate a test suite starting from Java source code using a search-based or random approach~\cite{Fraser2013IEEE, Panichella2018AutomatedTC}.
Several studies have uncovered challenges involving automatically generated tests~\cite{Palomba_2016, Palomba2016SBST, GRANO2019, Fraser_2013, Shamshiri_2015, Almasi_2017}, an important one being that 
generated tests are typically less readable than their human-written counterparts~\cite{grano2018empirical}. 
In this context, 
Almasi et al.~\cite{Almasi_2017} have 
observed that developers 
\begin{enumerate*}
\item find the test case scenario difficult to follow,
\item find the test data unclear, and 
\item have difficulties with the meaningfulness of generated assertions.
\end{enumerate*}

\subsection{Large Language Models}

Large Language Models are a subset of AI systems predominantly based upon the transformer architecture~\cite{vaswani2023attention}. These LLMs are trained on vast amounts of data, through which, they learn the underlying patterns inherent in texts, code, dialogue, etc., and are therefore capable of generating a (somewhat) relevant response given a prompt by the user~\cite{openai2023gpt4}. LLMs operate based on predicting the subsequent tokens in a sequence and reusing the extended sequence by running it through the model once again to predict the tokens to follow (referred to as autoregression). This process is continued up until a point in which either the maximum amount of required tokens is reached or a termination character is generated. 

Since their emergence, software engineers have utilized LLMs to enrich and simplify the development process~\cite{izadi2022codefill,izadi2024language,al2023extending}. In alignment with this, various open-source models, e.g., Code Llama~\cite{rozière2023code} and StarCoder~\cite{li2023starcoder}, and closed-source models, e.g., Codex~\cite{DBLP:journals/corr/abs-2107-03374}, and GPT4~\cite{openai2023gpt4}, 
have been trained and fine-tuned for this very purpose. 

The research community has recently investigated incorporating LLMs into the test-generation process. 
In particular, attempts have been made to evaluate the efficacy of utilizing existing LLMs for unit test generation~\cite{10329992,siddiq2024using}, and training specialized LLMs for test generation~\cite{10298372}. However, the understandability and usability of these (hybrid) tests remain unclear.
Additionally, methods such as CodaMosa~\cite{lemieux2023icse} and TestPilot~\cite{10329992}, respectively, propose the addition of LLMs to combat stalls in the search-based process and the full automation of the test generation process, which have proven to be useful. With all these additions, however, problems arise regarding the reliability, correctness, and complexity of incorporation given the non-deterministic nature of the results~\cite{ouyang2023llm,siddiq2024using}. 

Recent studies have pointed out that engineering good prompts is crucial for obtaining high-quality results~\cite{ouyang2023llm}. For instance, 
the usage of the Chain of Thought reasoning (CoT) method has been shown to provide major improvements in the zero-shot performance of models~\cite{wei2023chainofthought,li2023structured}. Furthermore, recent guidelines have been proposed that point toward the fact that including further information about the goal, context, and even the persona of the model can impact the quality of the results obtained~\cite{marvin2023prompt}. While these guidelines provide a good start to the process of constructing a quality prompt, in most cases the task remains an empirical process at heart~\cite{zamfirescu2023johnny}. 

\section{The \approach Approach}

\begin{figure}[!t]
\centering
\includegraphics[width=\linewidth]{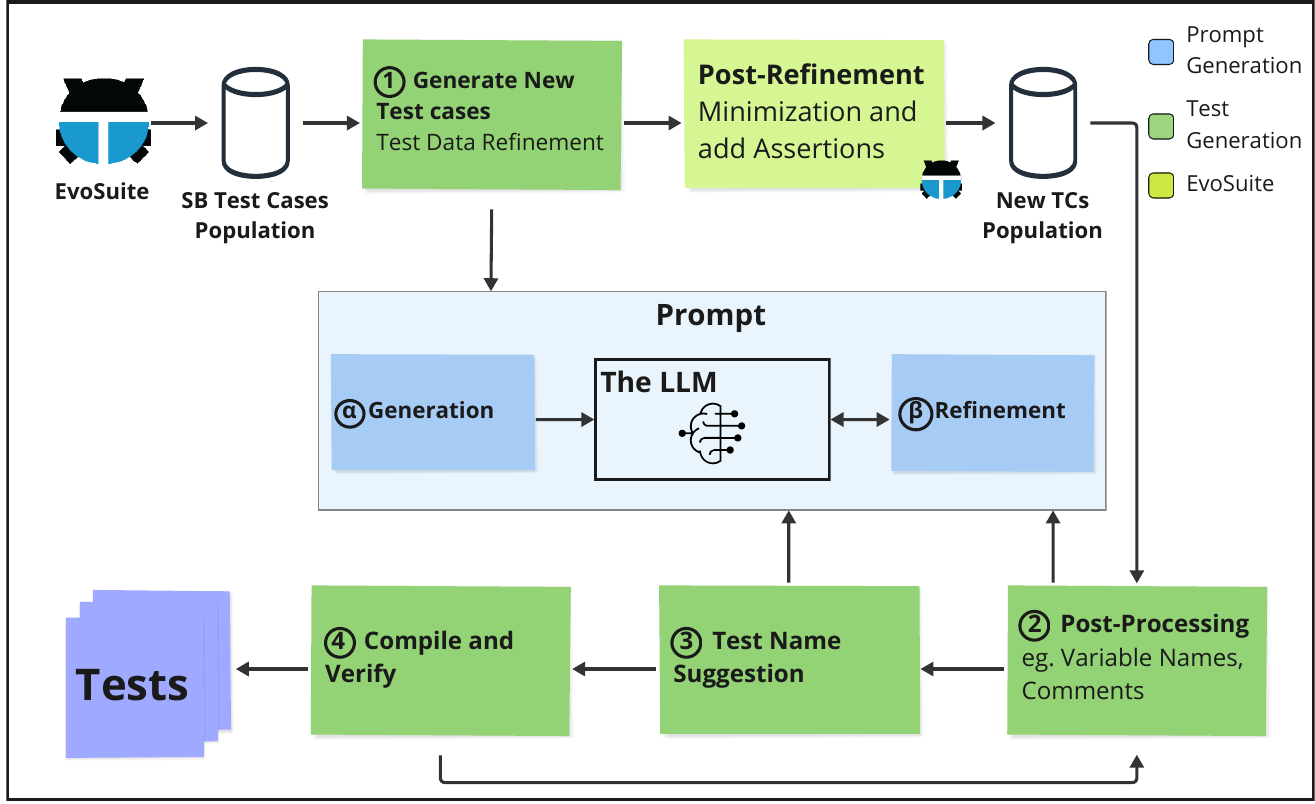}
\vspace{-7mm}
\caption{Overview of the UTGen approach}
\label{fig_overview}
\vspace{-4mm}
\end{figure}

Figure~\ref{fig_overview} provides an overview of our approach, named \approachNS. The core of our framework is a search-based approach in which we integrated an LLM in various stages of the test generation process (highlighted in green). 
We use EvoSuite~\cite{fraser2011evosuite} as the search-based test generation framework of choice, and we have developed additional functionalities that facilitate the integration of EvoSuite with LLMs (highlighted in blue).

The aim of our approach, UTGen, is to enhance the understandability of test cases by improving four key elements of generated tests: 1) providing context-rich test data, 2) incorporating informative comments, 3) using descriptive variable names, and 4) picking meaningful test names. These goals define the stages in our approach.

As a first step, after the genetic algorithm has ended and the test cases mature in the search-based process, our approach focuses on refining test data (\circled{1} in Figure~\ref{fig_overview}). UTGen uses an LLM to generate contextually relevant test data, unlike traditional search-based methods that often rely on random values. Following this refinement, the search process ends, and we transition to post-processing tasks. Here, EvoSuite minimizes the number of test cases in the test suite, shortens the length of individual tests, and adds assertions.

Once the test cases are fully formed, at stage \circled{2}, UTGen leverages an LLM to add descriptive comments and enhance variable names. 
In stage~\circled{3}, UTGen uses an LLM to suggest suitable names for the tests, reflecting the assertions and logic within. Finally, to ensure that test cases are compilable and stable after these enhancements, UTGen compiles them (stage~\circled{4}), and in case of compilation issues, the process iteratively revisits stage \circled{2} for adjustments.

We first explain the prompt engineering component 
and then describe our test generation process per stage.

\subsection{Prompt Generation}
The prompt component of \approach 
uses the \code{code-llama:7b-instruct} model from Meta~\cite{rozière2023code} 
as provided by Ollama\footnote{Ollama: https://ollama.com/}. 
We have designed \approach in such a way, 
that the Code-llama can easily be exchanged for another LLM. There are three stages within 
the \approach approach uses the prompt component:
\begin{enumerate*}
\item the refinement of test data,
\item the post-processing of tests, and
\item the naming of tests.
\end{enumerate*}
The general prompt component contains two distinct parts, namely \circledalpha{} which is responsible for generating the prompts provided to the LLM, and \circledbeta{} which manages the request and ensures the correctness of the returned response.

\begin{figure}[!t]
\centering
\includegraphics[width=\linewidth]{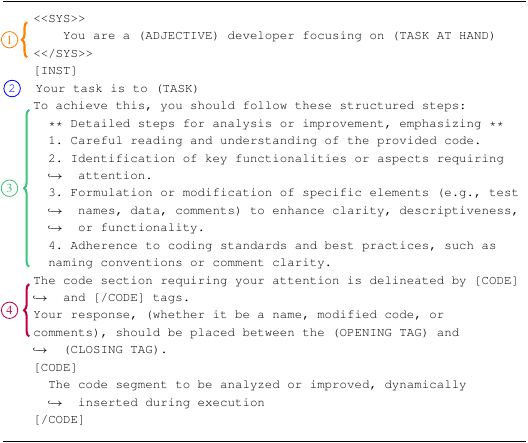}
\vspace{-6mm}
\caption{Prompt template for UTGen}
\label{listing:enhanced_prompt}
\vspace{-4mm}
\end{figure}

For each stage, we devised specialized prompts following guidelines from recent prompt engineering research~\cite{marvin2023prompt, wei2023chainofthought, li2023structured}.
As shown in Figure~\ref{listing:enhanced_prompt}, these guidelines emphasize the following: writing clear instructions with action words (as in~\circled{2}), adopting a persona for the model (as in~\circled{1}), allowing sufficient processing time through techniques like Chain of Thought (CoT) (as in~\circled{3}), standardizing input and output formats, and framing requests in a positive manner (as in~\circled{4}).
The starting point for each prompt resembles the one presented in Figure~\ref{listing:enhanced_prompt}.
As each model has its complexities, pitfalls, and preferred input format, no one-size-fits-all solution exists to prompt engineering, however, the guidelines set out above have guided us.
We have followed an iterative prompt engineering process in which each adjustment of the prompt was deliberated upon, before being accepted or rejected by the authors based on potential improvements in the results. 
An emerging pattern that we initially observed is that LLMs are incapable of always adhering to the output format
described for them. Therefore, we put guidelines in place to deal with such mismatches; as an example, we had to deal with cases where plain text was placed inside the code blocks, or when the intended delineation was not used by the LLM. 
Our replication package contains the final versions of the prompts that we engineered, in addition to other measures that were taken~\cite{replicationpackage}.

\subsection{Stage 1: Test Data Refinement}

In this stage, we focus on requesting contextualized test data from the LLM to increase the domain relevance of test data for a test scenario. 
We designed a parser that converts the LLM's responses into the structured format required by EvoSuite. 
 

The test data refinement stage should be considered as another iteration in the search process in which both new and original test cases coexist in the test population.
The refined test cases are capable of changing the logic of the original test, and they cover different parts of the method under test. 



\begin{figure}[!t]
\centering
\includegraphics[width=\linewidth]{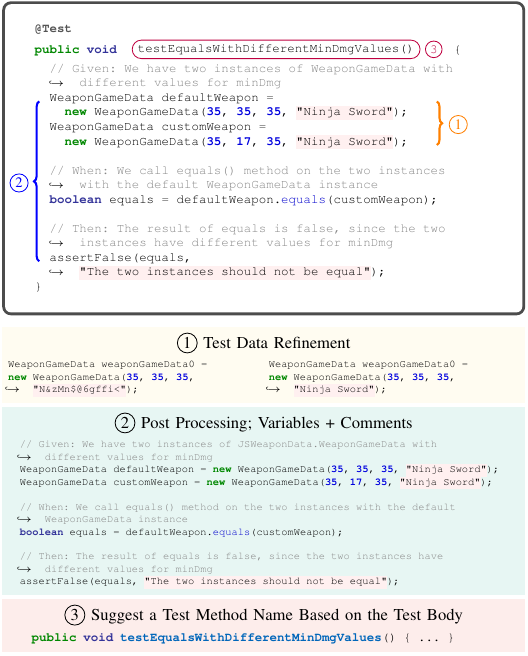}
\vspace{-8mm}
\caption{A simplified example of a test case enhanced by UTGen per step}
\label{fig_example}
\vspace{-4mm}
\end{figure}

An example of the refinement stage can be seen in stage~\circled{1} of Figure~\ref{fig_example}, with the original and enhanced test data shown side by side.
Based on the context, the LLM changes the fourth argument of the \texttt{WeaponGameData} constructor call from  
 \texttt{"N\&zMn\$@6gffi<"} into \texttt{"Ninja Sword"}, which is more meaningful in the context of \texttt{WeaponGameData}.

However, it is important to acknowledge certain limitations in the LLM's responses. Occasionally, the LLM may  hallucinate~\cite{fan2023large}, e.g., generate lines that deviate from the original test case, or alter the number of parameters in method invocations. 
To mitigate these inconsistencies, we designed our parser to substitute the erroneous line with the corresponding line from the original test case if a corresponding line exists for it in the original test case. In the absence of a corresponding line in the original test case, the parser will skip parsing these erroneous lines and continue parsing the remaining portions of the LLM-generated test cases.
This increases the chance that even test cases with omissions are valid for compilation.
For instance, if the LLM's response adds a non-existent statement like \code{weaponGameData0.increaseDmg(10)}, the parser skips this line and continues processing. Similarly, if the LLM alters a method's parameter count, like changing \code{weaponGameData0.getDmgBonus()} to \code{weaponGameData0.getDmgBonus(10)}, 
the parser uses the original method call with zero parameters.
These strategies ensure the parser extracts the maximum number of statements from the LLM responses, minimizing the need for re-prompting.

In post-refinement, EvoSuite optimizes the test case population and adds assertions to them. The optimization includes shortening test cases, and eliminating duplicated test cases from the population. The selection of which duplicate test case to keep and which to eliminate is directed by a secondary objective, which prioritizes selecting the test case that minimizes the total length of all test cases within the set of duplicates. 

\subsection{Stage 2: Post-Processing}\label{sec:approach:stage2}

In this stage, we make the final chosen test as understandable as possible by making various aspects of the code more understandable. \approach achieves this by adding descriptive comments, and making variable names more clear. 

After the post-refinement has finished, assertions are added to the test cases, and the test cases have reached maturity in terms of coverage, 
they are given to the LLM for improvement. The LLM is instructed to add comments (using the Given, When, Then convention --- seen as more understandable~\cite{Khorkov_2019}) and to exclusively change the naming of the variables but to let the data and logic untouched given that this could impact the intended behavior of a certain test.

To ensure maximal logical similarity between original and enhanced test cases, we use the CodeBLEU metric which effectively assesses syntactic and semantic similarities between two sequences~\cite{ren2020codebleu}.
We choose to control for similarity to increase the cohesion between generated and improved test cases as well as minimize the impact of LLM hallucinations. A CodeBLEU score below $0.5$ triggers a re-prompting process. 

We cap re-prompting at three iterations, as our findings suggest that this limit preserves logical coherence and still facilitates the improvement of tests.
If the LLM does not meet the threshold after three attempts, the prompt simplifies, removing comment structure constraints, and thus allowing deviation from the initial format. Should the LLM's response still not reach a satisfactory level after a total of six attempts, the original test case is retained. The value of three attempts per prompting strategy is also chosen to balance effectiveness and execution cost. 

Additionally, as previous literature has pointed out, results from LLMs can be non-deterministic given a non-modified temperature of the model being used, this can in turn lead to results diverging from the original tests, or tests that do not have correct syntax. To ensure consistency and reliability in the returned results, we employ a set of heuristic safeguards to facilitate the process of controlling for such anomalies. 
With each response from the LLM, we 
\begin{enumerate*} 
\item try to identify and remove common mistakes made by the LLM in the code, e.g., comments placed inside the code as plain text and not as comments,
\item attempt to correct any missing closing brackets in a piece of code,
\item validate code using CodeBLEU as previously described, and
\item check the syntactic correctness of the returned results with the parser generator tool ANTLR\footnote{ANTLR: https://www.antlr.org/}. 
\end{enumerate*} 

Furthermore, we re-prompt the LLM in the case when any of the previously described safeguards fail to improve the response, fail to achieve syntactic correctness, or have lower-than-threshold values for CodeBLEU.
We limit the amount of recursive calls that are made to not have a single improvement request stall the entire process. All the processes explained above relate to the component marked with \circledbeta{} in Figure~\ref{fig_overview}. 

An example of this step is shown in \circled{2} in Figure~\ref{fig_example}:
the comments in Given-when-then format are added, and the variable names are changed from \texttt{weaponGameData0} and \texttt{weaponGameData1} to the \texttt{defaultWeapon} and \texttt{customWeapon}, matching the logic of the test case. Also, the assertion message is added from the LLM response.


\subsection{Stage 3: Test Method Name Suggestion}

In this stage, UTGen gives the LLM the completed method body of the test, and it is asked to deduce a descriptive name. 
We chose to put this stage after the post-processing of the test method body because then the test case includes comments that increase the context for the LLM to generate a descriptive test method name. If another test case already has a similar test name, we re-prompt until it has a unique name. 

For instance, in \circled{3} in Figure~\ref{fig_example}, the LLM suggests \texttt{testEqualsWithDifferentMinDmgValues()}. This name 
reflects the test's functionality of examining the \texttt{equals} method across varying minimum damage values. In comparison, EvoSuite named this test \texttt{\small testCreatesWeaponGameData\-Taking6ArgumentsAnd\-CallsEquals3}.

\subsection{Stage 4: Compile and Verify}
After successfully navigating through the safeguards, it is still possible for a test case to fail to compile. Therefore, we compile all test cases, and a non-compiling test case undergoes a repeated cycle of post-processing and test method name suggestion, with a default post-processing budget of 2 iterations.

Compiling test cases are then assessed for their stability. A test case is considered unstable if it fails due to an exception unrelated to a JUnit assertion. 
All test cases that are both compilable and stable are saved.

\section{Experiment Setup}
In this section, we describe the methodology of evaluation of our approach. We investigate the following RQs:

\begin{description}
\item[\textbf{RQ$_{1}$}] \textit{Does \approach have the capability to generate effective unit tests by utilizing a combination of LLMs and SBST?}
\item[\textbf{RQ$_{2}$}] \textit{What is the impact of LLM-improved unit tests' understandability on the efficiency of bug fixing by developers?}
\item[\textbf{RQ$_{3}$}] \textit{Which elements of UTGen affect the understandability of the generated unit tests?}
\end{description}

We now discuss the evaluation strategies for RQ1 to RQ3.

\definecolor{Gray}{gray}{0.9}
\begin{table}[!t]
\centering
\caption{Demographics of Participants }
\vspace{-2mm}
\begin{smaller}
\begin{tabularx}{\columnwidth}{X X X X}
\toprule
\rowcolor{Gray} \textbf{Attendance} & Academia & Industry  & $\Sigma$ \\
\midrule
In Person & 17 & 4 & 21 \\
Remote & 3 & 8 & 11 \\
$\Sigma$ & 20 & 12 & 32\\
\end{tabularx}
\begin{tabularx}{\columnwidth}{X X X X X }
\midrule
\rowcolor{Gray} \textbf{Experience} & \multicolumn{2}{c|}{Academia} & \multicolumn{2}{c}{Industry}\\
\rowcolor{Gray} & In Java & \multicolumn{1}{l|}{In Testing} & In Java & In Testing \\
\midrule
0-2 years& 5& 9 &  0& 2\\
3-6 years& 11& 9&  8& 5\\
7-10 years& 2& 1 & 3& 2\\
$\ge$ 10 years& 2& 1 & 1& 3\\
\midrule
\end{tabularx}

\begin{tabularx}{\columnwidth}{l X l X l}
\rowcolor{Gray} \textbf{Affiliation} & \multicolumn{2}{c|}{Academia} & \multicolumn{2}{c}{Industry}  \\
\rowcolor{Gray}& Role & \multicolumn{1}{c|}{Number} & Role & Number\\
\midrule
&PhD Student& 9 (45\%) & Developer& 6 (50\%)\\
&MSc Student& 8 (40\%) & Senior Researcher&  3 (25\%)\\
&BSc Student &1 (5\%) &Scientific Dev.& 1 (8\%)\\
&Post Doc & 1 (5\%) &Team Lead& 2 (17\%)\\
&Scientific Dev. & 1 (5\%)\\
\bottomrule
\end{tabularx}

\end{smaller}
\label{table:participants}
\vspace{-4mm}
\end{table}

\subsection{Effectiveness Evaluation Setup (RQ1)}\label{sec:setup:RQ1}

We explore the effectiveness of UTGen on two axes: the compilability rate of LLM-improved test cases, and a comparison in coverage of baseline and UTGen test cases.


\subsubsection{Dataset}
We utilize the DynaMOSA dataset composed of $346$ non-trivial Java classes from $117$ open-source projects for RQ1~\cite{Panichella2018AutomatedTC}. 
The classes are selected from four different benchmarks, with the primary source being the 204 non-trivial classes of SF110~\cite{TOSEM_evaluation}. 






\subsubsection{Evaluation}
We evaluated UTGen using the EvoSuite framework as a baseline. 
We applied UTGen on a dataset and generated two types of test cases: original EvoSuite test cases and LLM-improved test cases. We then compare these two types of test cases by measuring 
\begin{enumerate*}
\item the number of LLM-improved test cases that compiled successfully, 
\item branch and instruction test coverage, and 
\item pass/fail rates.
\end{enumerate*}

\subsubsection{Parameter Configuration}
We decided to use the default configuration parameters for EvoSuite, which have been empirically shown to provide good results~\cite{arcuri2013parameter}. We did increase the test budget (max\_time) from 60 to 200 seconds, to ensure that the search algorithm has enough time to generate a test population that achieves reasonable coverage levels.

\subsection{Controlled Experiment (RQ2)}

We conducted a controlled experiment to assess the understandability of test cases in a real-world scenario, namely bug fixing~\cite{zellerBOOK}. 
This extends the work of Panichella et al., who investigated the impact of generating documentation for automatically generated tests in the context of bug fixing~\cite{panichella2016impact}.

The experiment involved 32 participants. 
The experimental group worked with UTGen test cases, while the control group was given EvoSuite test cases. We configured EvoSuite with \emph{coverage-based test naming}, which generates more readable test names than the default setting~\cite{Daka_2017}.

We examined two dependent variables in the experiment: \begin{enumerate*}
\item the number of fixed bugs, and 
\item time efficiency, measured as the time taken to fix the bugs.
\end{enumerate*}




\subsubsection{Participants}We recruited participants with academic and industrial backgrounds. Table~\ref{table:participants} presents their demographics. To engage academic participants, the experiment was advertised via the university's communication channels. Additionally, developers from an industrial partner were enlisted. Furthermore, all authors reached out to their professional networks of software engineers. We made sure to extend the invitation to individuals with experience in Java and testing.

\subsubsection{Objects}

\begin{table}[!t]
\centering
\caption{Java classes used for the controlled experiment}
\vspace{-2mm}
\begin{smaller}
\begin{tabular}{l l c c c}
\toprule
Project & Class & LOC & Methods & Branches\\
\midrule
caloriecount& Budget& 152& 21& 16\\
twfbplayer& JSWeaponData& 177& 19 & 44\\
\bottomrule
\end{tabular}
\end{smaller}
\label{table:classes}
\vspace{-5mm}
\end{table}

To design the bug-fixing assignments and compare experimental and control groups, it was essential to choose two projects that would offer a solid foundation for understanding the context of bug fixing.
To do so, we analyzed all classes within the SF110 dataset, gathered insights into the distribution of Lines of Code (LOC), which serves as an indicator of complexity~\cite{graylin2009cyclomatic}.
Using this data, we calculated the mean (\(\mu\)) and standard deviation (\(\sigma\)) for each distribution. We then identified all classes falling within the range of \(\mu \pm 0.1\sigma\) across all specified metrics. This process yielded a total of $15$ classes. Upon manual inspection of these classes, we selected two for consideration:
\begin{enumerate*}
\item \code{Budget}, which includes methods for calculating calories over intervals, and 
\item \code{JSWeaponData}, featuring methods related to Weapon Objects in a Java game.
\end{enumerate*}
Table~\ref{table:classes} provides details on the two classes. 
We inject four faults in each class, with each fault located in a different method under test. The injected bugs included replacement of arithmetic operations (2 bugs), statement deletion (1 bug), boolean relation replacement (2 bugs), and variable replacement (3 bugs). While the types of faults were similar across both classes, fixing the faults in the \texttt{Budget} class can be more challenging due to its detailed time calculations.


\subsubsection{Experimental Design}
Our experiment utilized a 2 × 2 factorial crossover design; it featured two periods and included a two-level blocking variable based on the object. In each period, subjects applied a different technique (treatment) to a different object (assignment).
We preferred the crossover design over a between-subject design due to the latter requiring a larger number of participants to achieve sufficient statistical power. 
The design of the experiment is detailed in Table~\ref{exp:tasks}, which outlines the four sequences used.  We followed the experimental design guidelines provided by Vegas et al.~\cite{vegas_crossover}.

To minimize learning effects, participants were given tasks involving different objects in each period. Additionally, to avoid any potential bias from optimal sequencing, we balanced the participants over the sequences in terms of the number of participants, and academic versus industry background.
For participants from academia, each sequence was executed~5 times, while for industrial participants it was executed~3 times. 

\subsubsection{Experimental Procedure}
The participants were able to execute the controlled experiment either in-person or remotely through videoconferencing. Before the actual experiment, we asked participants to fill in a pre-test questionnaire to gauge their experience.
One day before the experiment session we sent them
\begin{enumerate*}
\item a statement of consent,
\item instructions and materials for performing the experiment, including the two assignments,
\item a number indicating the sequence (see Table~\ref{exp:tasks}), and
\item a link to the online survey platform.
\end{enumerate*}
This advance preparation was necessary, because during the pilot evaluation we observed that receiving the projects just before the experiment led to additional preparation time, increasing the threat of tiredness. It could also lead to stress among participants if they encountered difficulties.


During the experiment, an examiner was continuously present to explain expectations and control any external factors that could affect the experiment, e.g., ensuring that participants did not use external sources to fix bugs.


In the experiment, we asked the participants to carry out two tasks; each task consisted of fixing four bugs in 30 minutes. We assume extending the time or having an unlimited window box could intensify the learning effect and introduce threats of tiredness/boredom.
If the participant indicated to have fixed all 4 bugs within the 30-minute time frame, the examiner double-checked this, and the participant could proceed to the next step. 
Each participant received two tasks:
\begin{enumerate*}
\item a task consisting of one Java class with a corresponding test class generated with UTGen, 
and 
\item a Java class with a corresponding test class generated by the baseline approach, i.e., EvoSuite.
\end{enumerate*}

\begin{table}[t!]
\centering
\caption{Experimental Design}
\label{exp:tasks}
\begin{smaller}
\begin{tabular}{c|c| l l | l l}
\toprule
\multirow{2}{*}{\#Seq} & \multirow{2}{*}{Order} & \multicolumn{2}{c}{Period 1}& \multicolumn{2}{c}{Period 2}\\
& & Object & Technique & Object & Technique\\
\midrule
I& U-E& Budget& UTGen& JSWeaponData& EvoSuite\\
II& E-U& Budget& Evosuite& JSWeaponData& UTGen\\
\midrule
III& U-E& JSWeaponData& UTGen& Budget& Evosuite\\
IV& E-U& JSWeaponData& Evosuite& Budget& UTGen\\
\bottomrule
\end{tabular}
\end{smaller}
\end{table}

\subsubsection{Pilot}
We engaged 4 participants (not part of the 32 participants) to pilot our experiment. 
After the pilot run, 
we changed the tasks from fixing 5 bugs in 20 minutes, to fixing 4 bugs in 30 minutes, and clarified the
expected behaviours through Javadoc documentation. 
We also narrowed the scope of the code, segregating it into \textit{definitely good} and \textit{possibly faulty} code sections. Thus ensuring that the assignments were feasible within the 30-minute time frame.
Finally, we improved the task descriptions, sending detailed instructions and an overview of the experiment to participants beforehand. 

\subsubsection{Analysis Method}

We conducted statistical tests to determine whether there was a significant difference between the number of bugs found and the time taken to fix bugs in LLM-improved test cases compared to baseline test cases. Due to our crossover design, we accounted for potential carryover effects, which required treating the data as dependent.
Therefore, nonparametric hypothesis tests for independent samples like the Wilcoxon Rank Sum test were not suitable~\cite{vegas_crossover}.

Instead, we employed mixed models for our analysis. Specifically, for each of our dependent variables:
\begin{enumerate}
    \item \emph{the number of fixed bugs}: this variable is discrete and bounded between~0--4, and we treated it as an ordinal variable. Consequently, we used Cumulative Link Mixed Models~\cite{christensen2018cumulative}, which are appropriate for this type of data.
    \item \emph{time efficiency}: this variable represents the time taken to fix bugs, and we used Generalized Linear Mixed Models with a Gamma distribution, which is suitable for time-related data~\cite{reaction_time}.
\end{enumerate}


We considered Technique, Object, Technique:Object, Order (confounded with carryover), and Period as fixed effects, and participants (\#id) as a random effect. The sequence effect is embedded within the variables Order and Technique:Object. We set the significance level at $0.05$ for both models.

Additionally, we examined whether factors such as participants' background, programming experience in Java and Testing, as well as whether the sessions were attended in-person or remotely, interacted with the technique on the number of fixed bugs. In these cases, we extended the mixed model by adding these factors to assess their interaction with the technique.

We also used Cohen’s d to measure the effect size ranging from very small ($d < 0.2$) to small ($0.2 \le d < 0.49$), medium ($0.5 \le d < 0.79$), and large ($d \ge 0.8$)~\cite{sullivan2012using}. 
 


\subsection{Post-Test Questionnaire (RQ3)}\label{sec:setup:RQ3}

We used the post-test questionnaire to obtain feedback from the participants of the controlled experiment on which aspects of UTGen affect the understandability of test cases (see Table~\ref{tab:questionnaire}).
We focused on gauging three aspects:
\begin{enumerate*}
\item participants' views on how the understandability of test cases impacts their bug-fixing effectiveness, 
\item their opinion on what factors in test code contribute to the understandability of generated test cases, and 
\item their ratings of the quality of these factors in test cases with and without the LLM-improved enhancements.
\end{enumerate*}

\subsubsection{{Questionnaire}}
In Q1, we ask participants to identify factors they believe to affect bug fixing effectiveness. Importantly, at this stage, the participants are unaware that the experiment focuses on the understandability of generated test cases, ensuring that their responses genuinely reflect their initial thoughts on bug fixing. In Q2, we query whether the participants think the clarity of generated test cases influences bug fixing.
Q3 and Q4 gauge which factors impact understandability most.
In Q5, we ask the participants to rate the understandability of the two tasks, using a Likert scale along with open-ended feedback.
Finally, in Q6 and Q7, we ask participants to rate specific elements such as comments, test data, test names, and variable naming in the test cases of both tasks
in terms of
\textit{completeness}, \textit{conciseness}, \textit{clarity}, and \textit{naturalness}, thus aiming for a detailed evaluation of different aspects of test case quality~\cite{roy2020deeptc, deljouyi2023generating, panichella2016impact}.

\begin{table}[!t]
\caption{Questionnaire overview}
\label{tab:questionnaire}
\centering
\setlength{\tabcolsep}{3pt}
\begin{smaller}
\begin{tabular}{lp{5.1cm}lc}
\toprule
\#& Title & Type of Question& Aspect\\
\midrule
Q1& In your opinion, what factors make finding bugs easier for you? & Open & 1 \\
Q2& Do you think the understandability of the test cases affects your bug fixing? & Likert & 1 \\
Q3& Prioritize the elements in helping understandability & Ranking & 2  \\
Q4& How important are the following elements in the understandability of the test case & Likert & 2  \\
Q5& How do you judge the understandability of the provided test case (Task 1 and 2) & Likert, Open & 3   \\
Q6& Evaluate how good you think the first task is in each item & Matrix Table, Open & 3   \\
Q7& Evaluate how good you think the second task is in each item & Matrix Table, Open & 3  \\
\bottomrule
\end{tabular}
\end{smaller}
\vspace{-4mm}
\end{table}

\subsubsection{{Analysis Method}}
For the open-ended questions, we sorted the data into categories using a card-sorting method and calculated the frequency of each category. Two authors independently reviewed the card-sorting process and achieved an 84\% agreement.
For the Likert-scale questions, 
we determined the mean value and percentage of each answer. For Q6 and Q7, 
we used the Wilcoxon Rank Sum test with an $\alpha$ of 0.05 because it did not follow a normal distribution (as determined by the Shapiro-Wilk test with a $p$-value $<< 0.01$). We used Cohen’s d effect size to determine the extent of the difference.

\section{Results}
In the following we
discuss the results per research question.

\subsection{RQ1: Effectiveness of Integrating LLMs and Search-Based Methods for Generating Unit Tests}

We define effectiveness as the capability of \approach to generate unit tests that are compilable and execute reliably, along with their ability to cover the classes under test. 
The success rate, defined as the proportion of generated tests that pass upon execution, reflects functional correctness. It is important to note that while all generated tests compile, the success rate pertains solely to their execution outcome.

\approach successfully generates a total of 8430 tests, with a pass rate of 73.27\%, while EvoSuite produces 8315 tests at a slightly higher pass rate of 79.01\%.
The heuristic safeguard 
described in Section~\ref{sec:approach:stage2} ensures the syntactic correctness and compilability of test cases, but also leads to 
27.52\% of the tests were categorized as ``\textit{enhancement stagnation}", i.e., the LLM could not improve the test case, or ``\textit{reverted}", i.e., we went back to the EvoSuite base test case, as the test case failed to compile. 
As such, these 27.52\% of test cases compile, but are not meaningfully affected by \approachNS. 

The origin of certain test cases not being meaningfully affected by \approach 
lies in the non-deterministic nature of LLMs.
As we have no guarantee that tests given to the LLM will compile upon improvement due to the possible hallucinations by the LLM, we employ several safeguards. 
While the safeguards explained in Section~\ref{sec:approach:stage2} do manage to catch a great portion of the tests that would not compile, some do fall through. Therefore, we perform a compilation check (\circled{4} in Figure~\ref{fig_overview}). 
If any (improved) test fails to compile, we revert back to an EvoSuite-generated test case. 



Out of the total 8430 tests generated by UTGen, 11.77\% are non-compiling and are thus reverted to the initial test case generated by EvoSuite. 
The remaining 15.75\% of tests are due to the stagnation of the enhancement process and the inability of the LLM to make a significant contribution. 

\begin{table}[!t]
\caption{Efficacy results of \approach-generated tests}
\setlength{\tabcolsep}{3pt}
\centering
\begin{smaller}
\begin{tabular}{p{0.15cm}p{2.5cm}p{2.1cm}p{1.9cm}}
\toprule
\# & 1. Pass/Failed & Test Count & Pass Rate\\
\midrule
1. & EvoSuite & 8315 & 79.01\%\\
2. & UTGen & 8430 & 73.27\%\\
\midrule
\# & 2. Improved Tests & Test Count & Percentage\\
\midrule
1. & Improved Tests & 6110 & 72.48\%\\
2. & Reverted Tests & 992 & 11.77\%\\
3. & Enhancment Stagnation & 1328 & 15.75\%\\
\midrule
\# & 3. Coverage & Instruction Coverage & Branch Coverage\\
\midrule
1. & EvoSuite & 25.03\% & 18.68\%\\
2. & UTGen & 24.43\% & 17.87\%\\
\bottomrule
\end{tabular}
\label{tab:test}
\vspace{-4mm}
\end{smaller}
\end{table}


Finally, 
from Table~\ref{tab:test} we observe that EvoSuite reaches slightly higher coverage compared to \approachNS{}: instruction coverage is 25.03\% compared to 24.43\%, while branch coverage is 18.68\% compared to 17.87\%. 
In a further investigation into the reason for this delta in coverage, we find that small changes in the post-processing step, e.g., changes in values of parameters, affect the overall coverage achieved.


\begin{figure*}[!t]
\centering
\includegraphics[width=0.85\linewidth]{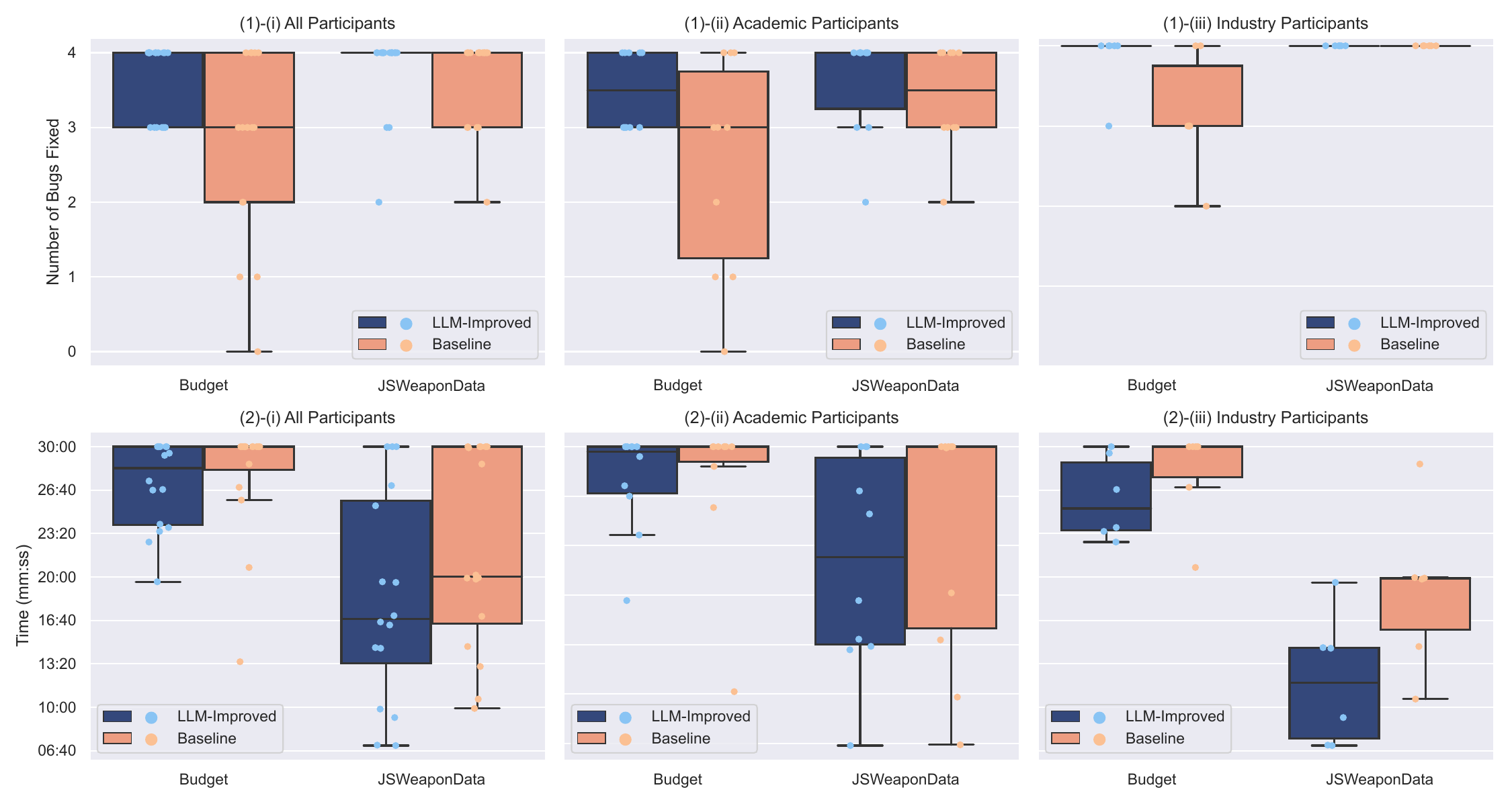}
\vspace{-3mm}
\caption{(1) Number of bugs and (2) Time taken for each different group: (i) All Participants (ii) Academic Participants (iii) Industry Participants}\label{fig:results_all}
\vspace{-2mm}
\end{figure*}



\begin{rqbox}
\textbf{RQ1} \textit{A total of 8430 tests are generated by UTGen, of which 72.48\% are improved, and 27.52\% are not due to reversion or stagnation. The coverage results are marginally comparable to the baseline.}
\end{rqbox}

\subsection{RQ2: The Impact on Bug Fixing}

Figure~\ref{fig:results_all} presents the results of the controlled experiment in terms of two dependent variables:
\begin{enumerate*}
    \item the number of bugs fixed, and 
    \item time efficiency, measured by the duration required to complete the tasks.
\end{enumerate*}  
The results are reported for respectively the entire population, the academic participants, and the industry participants.

For both objects, participants fixed more bugs in the task with the LLM-improved test cases compared to the baseline test cases.
For the \texttt{Budget} class the difference is more pronounced as the participants fixed a median of 4 bugs with LLM-improved test cases, compared to a median of 3 bugs fixed for baseline test cases. 
For the \texttt{JSWeaponData} class, the difference is marginal, as participants fixed a median of 4 bugs with either of the test cases. 
According to the tests of fixed effects  presented in Table~\ref{results:fixed_effects}, we observe in the fixed bugs column that both technique ($p = 0.024$) and object ($p = 0.025$) significantly influence the number of fixed bugs. This implies that using the LLM-Improved test cases significantly increases the likelihood of fixing more bugs. Similarly, when the object is JSWeaponData, the probability of fixing more bugs is also significantly higher. The result of Cohen’s d effect size for the treatment is medium at 0.59.

Regarding time efficiency, participants using LLM-improved test cases generally took less time to fix all bugs for both classes.

However, the differences in timing are not statistically significant for the technique ($p = 0.063$), with significance observed only for the object ($p = 0.031$).
The difference is more apparent in the \texttt{JSWeaponData} class, where the average time to fix all bugs was 18:22 for LLM-improved versus 22:06 for baseline test cases (20\% less time).
For the \texttt{Budget} class, the averages are closer: 27:06 for LLM-improved and 27:51 for baseline test cases. This is mainly due to a 30-minute cutoff, which limited the observable difference. 

Additionally, a post hoc analysis of Estimated Marginal Means involving a pairwise comparison of different technique levels for each specific object level indicates that in the \code{Budget} class, the treatment (LLM-Improved test cases) is significant ($p = 0.024$), whereas it is not significant in the \code{JSWeaponData} class ($p = 0.319$). The Cohen’s d effect size is large of $0.92$ for the treatment in the \code{Budget} class. We hypothesize that the statistically significant improvement in the number of bugs fixed in the \code{Budget} assignment, compared to \code{JSWeaponData}, is due to the greater complexity of scenarios and bugs in the \code{Budget} class. This complexity likely increases the demand for clearer and more understandable test cases.

Furthermore, neither the period ($p = 0.176$ and $p = 0.068$) nor the order ($p = 0.138$ and $p = 0.517$) significantly impact the number of fixed bugs and time efficiency. This indicates that there is no carryover effect between treatments.
The interaction between technique and object is not significant, suggesting that the effect of the technique on the number of bugs fixed and time efficiency does not depend on the object.
Additionally, our analysis found no significant interaction between the technique and co-factors such as participants’ backgrounds, experience in Java and testing, or whether they attended sessions remotely or in person ($p >> 0.05$).

\begin{table}[t!]
\centering
\caption{Tests of Fixed Effects}
\label{results:fixed_effects}
\vspace{-1mm}
\begin{smaller}
\begin{tabular}{l| c c | c c}
\toprule
\multirow{2}{*}{Source} & \multicolumn{2}{c}{Fixed Bugs}& \multicolumn{2}{c}{Time Efficiency}\\
& Estimate &  $Pr(>|z|)$  & Estimate &  $Pr(>|z|)$ \\
\midrule
Technique&  2.997& 0.024& -0.116& 0.063\\
Object& 2.903& 0.025& 0.133& 0.031\\
Technique: Object& 0.951& 0.401& 0.088& 0.408\\
\midrule
Order& 1.588& 0.138& -0.069& 0.517\\
Period& 0.951& 0.176& -0.114& 0.068\\
\bottomrule
\end{tabular}
\end{smaller}
\vspace{-3mm}
\end{table}


Finally, in terms of the influence of the background of our participants, we observe that both population groups show better performance when using LLM-improved test cases compared to baseline test cases in terms of both number of bugs fixed and time taken to fix bugs.  
We observe that academic participants seem to benefit more from the LLM-improved test cases in aiding bug fixing. 
For industrial participants on the other hand, the time-saving gain is more pronounced. Figure~\ref{fig:results_all} provides a more detailed overview.

\begin{rqbox}
\textbf{RQ2}
\textit{In our experiment, using LLM-Improved tests significantly increases the likelihood of fixing more bugs.}

\end{rqbox}



\subsection{RQ3: The effects of different elements of UTGen on understandability}

The results of the post-test questionnaire show three aspects: \begin{enumerate*}
\item the participants' views on how the understandability of test cases impacts their bug-fixing effectiveness, 
\item their opinion on what factors in test code contribute to the understandability 
and 
\item their ratings of the quality of elements in test cases with and without the LLM-improved enhancements.
\end{enumerate*}


\paragraph*{Aspect 1: How understandability of test cases impacts bug-fixing} We answer the first aspect through the responses to 
Questions~1 and~2 in the survey. We have observed that participants find a well-written test suite important for bug fixing: 
they frequently highlighted (14 mentions) the importance of descriptive and clear test names, appropriate use of assertions, and well-chosen test data in test suites. 
This aspect was prioritized over other factors like high-quality production code
(10 mentions).
We also take note of the overall (strong) agreement that test case understandability is important in
in the context of bug fixing, as indicated by a median score of 4 out 5 (Q2 in Table~\ref{tab:rq3}).

\begin{table}[!t]
\caption{Participants' responses to the questionnaire}
\vspace{-1mm}
\setlength{\tabcolsep}{4pt}
\centering
\begin{smaller}
\begin{tabular}{p{2.1cm}|p{.96cm}p{.96cm}p{.96cm}p{.96cm}p{.96cm}}
\toprule
\multirow{2}{2.1cm}{Q2. The effect of understandability on bug fixing} & Strongly disagree & Disagree & Neither agree nor disagree & Agree & Strongly agree\\
\cmidrule(lr){2-6}
& 6.2\% & 9.3\% & 6.2\% & 34.4\% & \textbf{43.7\%} \\
\midrule
\multicolumn{2}{p{3.03cm}|}{Q3. Prioritize the elements in helping understandability} & Rank 4 & Rank 3 & Rank 2& Rank 1 \\
\cmidrule(lr){3-6}
\multicolumn{2}{p{3.03cm}|}{~~1. Comment} & 9.3\% & 25\% & 31.2\% & \textbf{34.3\%}\\
\multicolumn{2}{p{3.03cm}|}{~~2. Test Name} & 34.3\%& 9.3\% & 15.6\% & \textbf{40.6\%}\\
\multicolumn{2}{p{3.03cm}|}{~~3. Variable Naming} & 21\% & \textbf{34.3\%} & \textbf{34.3\%} & 9.3\% \\
\multicolumn{2}{p{3.03cm}|}{~~4. Test Data} & \textbf{34.3\%} & 31.2\% & 18.7\% & 15.6\% \\
\midrule
Q4. How important are the elements in the understandability & Not important & Slightly important & Moderately important & Very important & Extremely important \\
\cmidrule(lr){2-6}
~~1. Comment& 6.2\%& 15.6\%& 21.8\%& \textbf{34.3\%}& 21.8\% \\
~~2. Test Name& 6.2\%& 21.8\%& 21.8\%& 12.5\%& \textbf{34.3\%} \\
~~3. Variable Naming& 3.1\% & 12.5\% & \textbf{34.3\%} & 31.2\% & 18.7\%\\
~~4. Test Data& 0\% & 12.5\%& 28.1\% & \textbf{43.7\%}& 15.6\% \\
\midrule
Q5. The quality of test cases & Very low & Low & Moderate & High & Very high \\
\cmidrule(lr){2-6}
~~LLM-improved& 3.13\%& 6.25\%& 25.0\%& 50.0\%& 15.63\% \\
~~Baseline& 6.25\%& 18.75\%& 25.0\%& 43.75\%& 6.25\% \\
\bottomrule
\end{tabular}
\label{tab:rq3}
\vspace{-5mm}
\end{smaller}
\end{table}

\paragraph*{Aspect 2: What factors in test code contribute to  understandability}
We have analyzed the participants' responses to Questions 3 and 4, where they ranked and scored the importance of elements. 
From Table~\ref{tab:rq3} we observe that participants give more importance to comments and test names, than to variable naming and test data. Specifically, 34.3\% of the participants ranked comments as most important, while 40.6\% gave priority to test names in Question 3.


\paragraph*{Aspect 3: The quality of factors in test cases with and without LLM enhancements} 

In Q5 of Table~\ref{tab:rq3}, we see that participants rate the understandability of LLM-improved tests somewhat better when compared to the baseline test cases. 


We asked participants in Q6 and Q7 to evaluate an LLM-improved and baseline test case of the assignments on different criteria and specifically per test element. 
These criteria comprised \textit{completeness}, \textit{conciseness}, \textit{clarity}, and \textit{naturalness}.

\begin{figure*}[!h]
\centering
\includegraphics[width=0.85\linewidth]{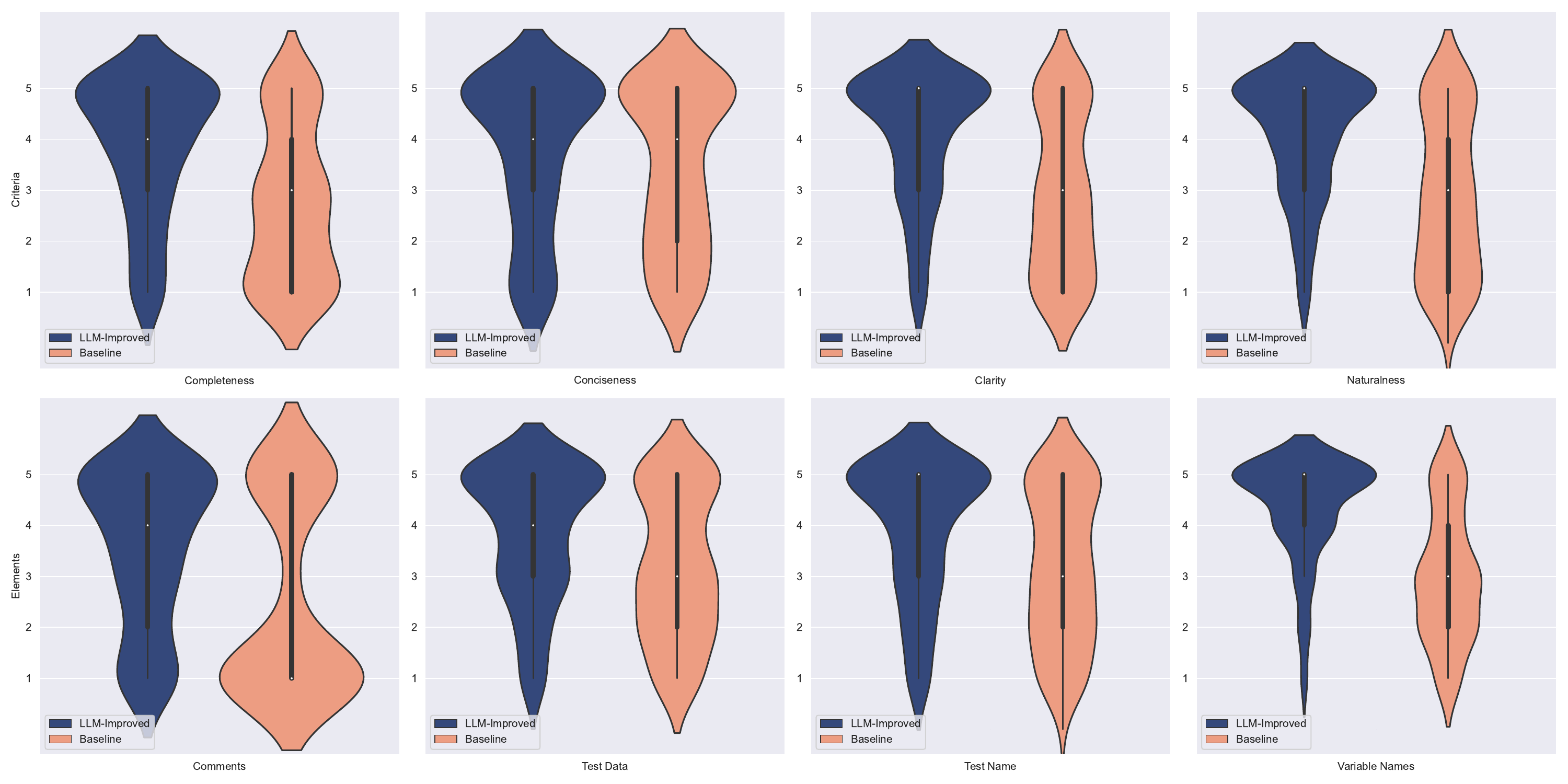}
\vspace{-5mm}
\caption{The results of Q6 and Q7 which test cases were rated in terms of criteria (first row) and test elements (second row)}\label{fig:results_criteria}
\vspace{-4mm}
\end{figure*}

Figure~\ref{fig:results_criteria} shows the results of Q6 and Q7. The results indicate that LLM-improved test cases are consistently rated higher compared to baseline test cases for each of the criteria (first row). 
The Wilcoxon test confirms this statistically significant difference ($p$-value $< 0.05$) for all criteria. The effect size for conciseness was small, while it was large for completeness, naturalness, and clarity.
Notably, in the open-ended responses, some participants mentioned that some comments in LLM-improved test cases were too general and added little value. The respondents did appreciate the Given-When-Then-structured comments. 


When we zoom into the test elements, we see improvements in all areas for LLM-improved test cases: comments, test data, test name, and variable naming. The Wilcoxon test for all of these elements is statistically significant with a $p$-value $<< 0.05$.
The effect size for comments, test data, and test names is medium, while very large for variable naming ($d > 1.2$). 

Through the analysis of the open-ended responses to Questions 5--7, we found that the complexity of a test case has an impact on the necessity of comments. For simpler test cases, using a Given-When-Then (Arrange/Act/Assert) structure is often sufficient. However, for more complex cases, more detailed comments are needed to ensure optimal comprehension. Overall, participants mentioned this point 18 times, with one participant stating  ``The test code lines are straightforward, so comments are unnecessary.''
Similarly, for simpler test cases, the quality of variable naming was less of a concern: participants mentioned this factor only 4 times when rating a short baseline test case.




\begin{rqbox}
\textbf{RQ3}
\textit{Comments, test names, variable names, and test data are improved compared to the baseline. Specifically, participants highlighted improved conciseness, clarity, and naturalness in these test elements.}
\end{rqbox}

\section{Discussion}
In this section, we discuss our results, their implications, and threats to the validity of our study.


\subsection{Revisiting the Research Questions}

\noindent\textbf{RQ1:} \emph{Does \approach have the capability to generate effective unit tests by utilizing a combination of LLMs and SBST?}
When we compare the effectiveness of our LLM-inspired \approach approach and EvoSuite, we observe that \approach generates test cases that have relatively similar structural coverage. 
However, we also noticed a phenomenon that we term \emph{enhancement stagnation}, which occurs when the LLM is not able to improve the test case, even when re-prompting multiple times. We analyzed this situation and found indications that this stagnation is correlated with high complexity. In this context, we define complexity at the level of the class under test to be: 
\begin{enumerate*}
\item methods having a high number of parameters, and
\item methods being tightly coupled, i.e., many method calls between objects or within an object.
\end{enumerate*}
While generally adding more relevant context can help an LLM, highly complex projects can overwhelm LLMs due to lengthy input codes and insufficient contextual information, thus hindering the enhancement process during post-processing. To overcome this, we propose to incorporate Retrieval Augmented Generation (RAG) techniques. We hypothesize that these enhancements can reduce occurrences of Enhancement Stagnation as it has resolved similar stagnation issues in other domains~\cite{parvez2021retrieval, liu2021retrievalaugmented}. RAG involves enhancing LLMs by dynamically integrating 
knowledge from databases, knowledge graphs, or the internet in real time into the generation process to provide contextually richer and more accurate responses.

\smallskip
\noindent\textbf{RQ2:} \emph{What is the impact of LLM-improved unit tests' understandability on the efficiency of bug fixing by developers?}
From the results of the controlled experiment, we see indications that the LLM-based enhancements brought to the generated unit tests improve their understandability in the bug-fixing scenario. Specifically, the experimental group outperformed the control group by fixing up to 33\% more bugs and completing tasks up to 20\% faster.
Our experiment consisted of two assignments involving respectively the \code{Budget} and \code{JSWeaponData} classes. While we observed statistically significant improvements for the \code{Budget} assignment, the other assignment did not reach statistical significance. Since the \code{Budget} class is comprised of more complex scenarios and bugs, we hypothesize that the complexity of a test scenario increases the need for understandable test cases. This hypothesis was anecdotally confirmed by participants in the post-test questionnaire. 

\smallskip
\noindent\textbf{RQ3:} \emph{Which elements of UTGen affect the understandability of the generated unit tests?}
Through the post-test questionnaire, we captured that participants think that LLM-improved test cases are showing improvements in terms of comments, test names, test data, and variable names when compared to baseline test cases. At a higher level, participants also rated completeness, conciseness, clarity, and naturalness as better. However, feedback from open-ended questions highlights that comments should be more precise and informative. Similarly, some participants also highlighted that simple test methods might not require (extensive) comments. Upon reflecting on this feedback, we hypothesize that generically trained LLMs, while generally robust, might lack task-specific data to effectively assist in creating comments.


\subsection{Implications}
Our study’s results have an important implication for researchers and tool builders. 
In particular, our study indicates that a generally trained LLM can already instigate a considerable improvement in the understandability of search-based generated test cases. However, our results also show that test case comments should be more detailed in some cases, while seeming superfluous in other situations. Therefore, we see potential in creating specifically-trained LLMs for particular software engineering tasks, but equally in customizing LLM responses to individual software engineers. 

\subsection{Threats to Validity}
\textbf{Construct Validity}. Threats to construct validity relate to the setup of our study. We conducted the study either in person or remotely, with an examiner present. To control for factors other than the codebase, we ensured a consistent setup for all participants and limited the choice of IDE to IntelliJ, providing uniform capabilities. However, this approach may disadvantage participants having experience with other IDEs, potentially affecting their performance. 

\textbf{Internal Validity}. To mitigate threats to internal validity, we did not reveal the tool names in our experiment and questionnaire. To prevent bias in selecting classes for the assignments, we followed a systematic selection process to strengthen the methodological integrity.
To compensate for a learning effect, 
we created four different 
sequences of the experimental design. 
Using mixed models, we found that period and carryover effects were not statistically significant, indicating they do not pose major threats to the study’s validity.

\textbf{External Validity}.
The classes that we use to determine the efficacy of test generation in RQ1 are a potential threat to the generalization of our results. To address this, we used a dataset of 346 classes from 117 open-source Java projects that form a representative sample and were previously used in software testing studies~\cite{Panichella2018AutomatedTC, CAMPOS2018207}.  
We limited the controlled experiment in RQ2 to two Java classes. To ensure their representativeness, we carefully selected them from the SF110 dataset containing real-world classes, and taking the average LOC of that entire dataset into consideration to select ``average classes''.
Future work will explore more complex classes. 
In order to mitigate potential imbalance between the experimental and control groups, we carefully balanced participants over both groups in terms of experience and background. 

\section{Related Work}

\subsection{Improving the Understandability of Test Cases}
Panichella et al.~\cite{panichella2016impact} introduced TestDescriber, which generates test case summaries that describe the intent of a generated unit test; they established that these summaries enable software engineers to resolve bugs more quickly. 
Similarly, Roy et al.~\cite{roy2020deeptc} developed DeepTC-Enhancer, leveraging deep learning to produce method-level summaries for test cases. Both efforts highlight the value of summarizing test cases. 
In contrast, \approach generates 
detailed comments within the test cases themselves 
and provides a narrative of the test scenario.

Zhang et al.~\cite{Zhang_2016} introduced an NLP technique for automatically generating descriptive unit test names.
Daka et al.~\cite{Daka_2017} applied coverage criteria for naming
automatically generated unit tests, 
while Roy et al.~\cite{roy2020deeptc} created DeepTC-Enhancer by employing deep learning to rename identifiers in test cases to improve readability.
Unlike these methods that rely on traditional NLP techniques, 
\approach utilizes LLMs to suggest identifiers that fit the test scenario's context. 

Afshan et al.~\cite{Afshan_2013} enhanced the readability of inputs by combining natural language models with search-based test generation. 
Deljouyi et al.~\cite{deljouyi2023generating} proposed an approach that generates understandable test cases with meaningful data through end-to-end test scenario carving. Baudry et al.~\cite{baudry2024generative} developed a test data generator using LLMs to produce realistic, domain-specific constraints. Our method is similar to Baudry et al.'s, but we focus on search-based unit test generation.

\subsection{Generating Test Cases by LLM}
Despite the progress in LLM-based test generation, to the best of our knowledge, no study has focused on enhancing unit test case understandability through the integration of search-based methods and LLMs. Research in this field shows considerable variability in methods and outcomes. 
Siddiq et al. generated tests using LLMs and reported 2\% coverage on the SF110 dataset~\cite{siddiq2024using}. In contrast, Sch\"{a}fer et al.'s~\cite{10329992} TestPilot for JavaScript achieved 70\% statement-level coverage on relatively small systems. 
Alshahwan et al. aimed to improve human-written tests by LLMs and submit them for human review~\cite{alshahwan2024automated}. 
Meanwhile, Lemieux et al. explored overcoming coverage stalls in SBST with LLMs~\cite{lemieux2023icse}, and Moradi et al. investigated mutation testing with LLMs~\cite{dakhel2023effective}. Steenhoek et al. improved test generation by minimizing test smells through reinforcement learning~\cite{steenhoek2023reinforcement}.
Unlike the aforementioned studies, \approach focuses on enhancing understandability through integrating LLMs in the SBST process. Notably, UTGen achieved 17.87\% branch coverage, surpassing the pure LLM approach by Siddiq et al.~\cite{siddiq2024using}. 

\section{Conclusion}
Recent research has suggested that the understandability of test cases is a key factor to optimize in the context of automated test generation~\cite{Almasi_2017}. Therefore, in this paper, we introduce the \approach approach that incorporates a Large Language Model (LLM) into the Search-Based Software Testing (SBST) process. In doing so, \approach aims to improve the understandability by providing context-rich test data, informative comments, descriptive variables, and meaningful test names.

We first evaluated \approachNS{}'s test generation effectiveness on 346 non-trivial Java classes, observing that \approach successfully enhanced 72.48\% of the test cases, and slightly decreased coverage compared to EvoSuite-generated tests (RQ1). 
We then performed a controlled experiment with 32 participants from industry and academia; we observed that test cases generated by \approach facilitated easier bug-fixing with participants fixing up to 33\% more bugs and doing so up to 20\% faster (RQ2). 
Feedback from participants in the post-test questionnaire indicated a significant improvement in test case completeness, conciseness, clarity, and naturalness (RQ3).


In future work, we aim to explore optimization strategies, such as Retrieval Augmented Generation (RAG), to enhance prompt efficiency and minimize the need for re-prompting. 
Furthermore, we plan to refine our approach by creating customized fine-tuned LLMs specifically for test generation. These customized LLMs would replace the publicly-available pre-trained LLM that we currently use.

\section*{Acknowledgement}
 This research was partially funded by by the Dutch science foundation NWO through the Vici ``TestShift'' grant (No. VI.C.182.032). 

\bibliographystyle{IEEEtran}
\bibliography{main}
\end{document}